\documentclass[fleqn,twoside]{article}
\usepackage[headings]{espcrc2}

\usepackage{epsfig}
\usepackage{graphicx}
\usepackage{rotating}
\usepackage{fleqn}
\usepackage{amsmath}
\usepackage{espcrc2}

\newcommand{\thtp}{$\Theta_5(1540)^+$}
\newcommand{\ximm}{$\Xi_5(1860)^{--}$}
\newcommand{\xiz}{$\Xi_5(1860)^0$}
\newcommand{\pK}{$p K_S^0$}

\def\etal {{\it et~al.,}}
\input babarsym

\title{\vspace{-2.9cm}
\begin{flushleft}\small{Presented at QCD 05: High Energy Physics International Conference
       in Quantum Chromodynamics, Montpellier, France, 4-8 Jul 2005.}\\
       \vskip 5pt
      \small{SLAC-PUB-11527}
       \end{flushleft}         
       \vskip 20pt
       Search for Pentaquarks at \babar}

\author{K. G\"otzen\address{Institut f\"ur Experimentalphysik I, Ruhr-Universit\"at Bochum, \\
	44780 Bochum, Germany
	(on behalf of the \babar\ Collaboration)}
}

\runtitle{Search for Pentaquarks at \babar}
\runauthor{K. G\"otzen}

\begin{document}

\begin{abstract}
The results of searches for the strange pentaquark states, \thtp, \ximm\ and
  \xiz\ in data recorded by the \babar\ experiment are  presented.  We
  search for these three states inclusively in 123.4 fb$^{-1}$ of $e^+ e^-$ annihilation
  data produced at the PEP-II asymmetric storage rings;  we find no evidence
  for their production in any physics process, and set limits on their
  production rates that are well below the measured rates for conventional
  baryons.  We also search for \thtp produced in interactions of electrons or
  hadrons in the material of the inner part of the detector.  No evidence
  for this state is found in a sample with much higher statistics than
  similar electroproduction experiments that claim a signal.\end{abstract}

\maketitle

\section{Introduction}
Several experiments have recently claimed observations \cite{hicks} of narrow baryonic resonances with
exotic quantum numbers whose interpretations as quark bound states require a minimum content of five quarks; these states are denoted as \thtp\ ($uudd\bar{s}$), \ximm\ ($ddss\bar{u}$) and its corresponding  partner \xiz. On the other hand, null results have also been published and have now outnumbered the positive claims. The interpretation and the comparison of the various experimental results is complicated by the problem of comparing different production mechanisms and energy ranges. Therefore it is of interest to perform high statistics and high resolution searches which encompass different production processes.

This has been done by \babar\ by measuring upper limits for the production of pentaquark states in $e^+ e^-$ annihilations on the one hand and  on the other hand performing a study of the electro- and hadro-prodution of the pentaquark state \thtp\ in the material of the inner part of the \babar\ detector.    

\section{The \babar\ Detector}
The \babar\ experiment is taking data at the PEP-II $e^+ e^-$ collider at center of mass
energy 10.58 GeV. The \babar\ detector is described in detail elsewhere \cite{nim}. Charged particle track parameters are measured by a five-layer double-sided silicon vertex tracker and a 40-layer drift chamber located in a 1.5-T magnetic field. Charged particle  identification is achieved with an internally reflecting ring image Cherenkov detector (DIRC), and by making use of specific energy loss $(dE/dx)$ measured in the tracking devices. Photons and neutral pions are detected with an electromagnetic calorimeter consisting of 6580 CsI(Tl) crystals. An intrumented flux return provides muon identification and detection of long-lived neutral hadrons.

\section{Inclusive search in \boldmath$e^+ e^-$ annihilations}
We search for the inclusive production of pentaquark states in $e^+ e^- \rightarrow P X$ reactions with any final state $X$ recoiling against the pentaquark candidate $P$. The analysis is based on 123.4 fb$^{-1}$ of data recorded at or slightly below the $\Upsilon(4S)$-resonance \cite{hep0502004}.
Investigations of several of the possible strange antidecuplet members usually considered as pentaquark candidates have been performed \cite{hep0408064},  and in particular the final results for the states \thtp, \ximm\ and \xiz\ are reported on here.

\begin{figure}[]
\begin{center}
\includegraphics[width=0.45\textwidth]{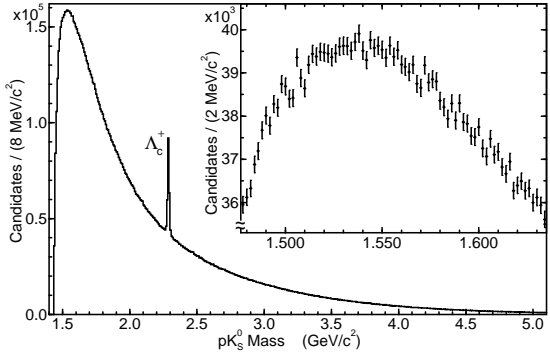}
\caption{The distribution of the $pK_S^0$ invariant mass for combinations satisfying the criteria described in the text. The same data are plotted for the full kinematically-allowed $pK_S^0$ mass range and, in the inset, with statistical uncertainties and a suppressed zero on the vertical scale, for the mass range in which the \thtp\ has been reported.}
\label{fig:pksmass}
\end{center}
\end{figure}

We reconstruct \thtp\ in the \pK\ decay mode, where $K_S^0\rightarrow \pi^+ \pi^-$. A sample of $K_S^0$ candidates is obtained from all pairs of oppositely-charged tracks identified as pions, that are consistent with a common vertex, and yield an invariant mass and momentum vector in accord with $K_S^0$ production from the $e^+ e^-$ collision axis. The $p$ and $\bar{p}$ candidates are identified by making use of the $dE/dx$ information measured with the tracking system and the Cherenkov angle $\theta_C$ determined with DIRC. Fig. \ref{fig:pksmass} shows the distribution of the \pK\ invariant-mass. No enhancement is seen in the region of the reported \thtp\ mass (inset in Fig.\ref{fig:pksmass}), whereas a clear peak at 2285 \mevcc with a mass resolution of about 6 \mevcc containing 98,000 entries is visible, originating from $\Lambda_c \rightarrow p K_S^0$ decays. To be independent of any production mechanism model leading to a specific \pK\ momentum spectrum in CM frame ($p^\ast$) different from that of the background, we split the data into subsamples according to the value of $p^\ast$, where each $p^\ast$ interval is 500 \mevc wide. No evidence for a signal is found in any of these subsamples.

We quantify this null result for a \thtp\ mass of 1540 \mevcc by fitting a convolution of a double-Gaussian (HWHM $\approx$ 2 \mevcc)and a $P$-wave Breit-Wigner as the signal lineshape and a background polynomial to the invariant-mass distribution of each subsample to determine the differential production cross section $d\sigma/dp^\ast$ from the fitted yields. 

Since the intrinsic width of the \thtp\ has not been measured until now, we use $\Gamma=1 \mev$ (for a narrow \thtp) and $\Gamma=8 \mev$ (best upper limit) and quote results for each assumed width. We determine the 95\% confidence level upper limit for the number of produced pentaquarks per $e^+ e^- \rightarrow hadrons$ event and compare it to the known production rates of conventional baryons, assuming \BR(\thtp$\rightarrow p K_S^0$) = 25\%.
The extracted numbers $5.0\times 10^{-5}$/event ($\Gamma=1\mev$) and $11\times 10^{-5}$/event ($\Gamma=8\mev$) are between 8 and 15 times lower than expected for conventional baryons. The corresponding cross section upper limits are 171 fb and 363 fb.

\begin{figure}[htb]
\begin{center}
\epsfig{file=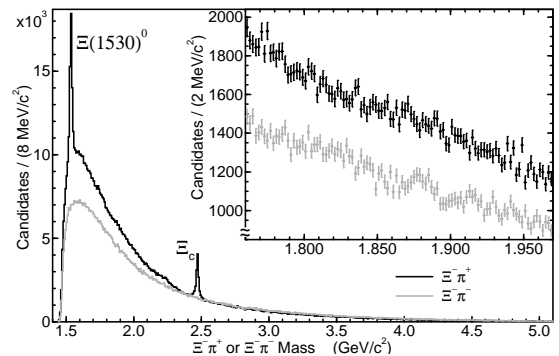, clip=, width=0.45\textwidth}
\caption{The $\Xi^-\pi^+$ (black) and $\Xi^-\pi^-$ (grey) invariant mass distributions for combinations satisfying the criteria described in the text. The same data are plotted for the full kinematically-allowed $\Xi^-\pi^\pm$ range and, in the inset, with statistical uncertainties and a suppressed zero on the vertical scale, for the mass range in which the \ximm\ and \xiz\ have been reported.}
\label{fig:xipimass}
\end{center}
\end{figure}

We search, as well, for the reported \ximm\ and \xiz\ states decaying into a $\Xi^-$ and a charged pion, where $\Xi^- \rightarrow \Lambda^0 \pi^-$ and $\Lambda^0\rightarrow p\pi^-$. The $\Lambda^0$ candidates are selected from all pairs of oppositely-charged tracks satisfying  proton and pion identification requirements, and are consistent with a common origin.
Fig. \ref{fig:xipimass} shows the invariant-mass distributions for $\Xi^- \pi^-$ and for $\Xi^- \pi^+$ combinations. In the $\Xi^-\pi^+$ mass spectrum, we see clear signals for the $\Xi(1530)^0$ and $\Xi_c(2470)$ baryons with 24,000 and 8,000 entries respectively, but no other structure is visible. There are no visible narrow  structures in the $\Xi^- \pi^-$ mass spectrum; the mass resolution at $\approx$ 1.86 \gevcc is 8 \mevcc.
As in the previous analysis we assume two different  intrinsic widths of this pentaquark state, namely $\Gamma = 1 \mev$ (narrow) and $\Gamma = 18 \mev$ (resolution of the signal claimed by NA49, considered as upper limit) to determine the 95\% confidence upper limit of the production rate in $e^+ e^-$ interactions. The results $0.74\times 10^{-5}$/event (for narrow width) and $1.1\times 10^{-5}$/event are 4-6 times lower than those for conventional baryons. The corresponding cross section upper limits in this case are 25 fb and 36 fb, respectively.

\section{Search for \boldmath\thtp\ in electro- and hadro-production}
Most of the positive evidence for exotic pentaquark states has been found in experiments based on photo-, electro- or hadro-production reactions on nuclear targets. Therefore a second analysis is performed by \babar\, which extends the search for \thtp\ to the interactions of secondary hadrons (tracks of every type) and beam-halo electrons and positrons in the material of the inner part of  the \babar\ detector, leading to inclusive production of the $p K_S^0$ system.

To reconstruct the $p K_S^0$ candidates, protons are identified by evaluating the specific energy loss information, $dE/dx$, measured with the tracking system. $K_S^0$ candidates are selected from all pairs of oppositely-charged tracks, which have a maximum distance of closest approach (DOCA) of 3 mm, a minimum flight length of 2 mm, and whose kinematic fit resulted in a chi-square probability $P(\chi^2)> 0.001$.
A candidate ($K_S^0$, $p$) vertex is defined as the mid-point of the DOCA (required to be $<$ 3 mm) of the $K_S^0$ flight path and the $p$ track. The radius of this vertex with respect to the collision axis had to be $>$ 2 cm; this limit is inside the beampipe, but well away from the collision axis.

The candidates are shown to reproduce the detector geometry to a high degree of accuracy giving confidence, that these events are due to interactions in the detector material. However the inclusive $p K_S^0$ invariant-mass distribution shows no evidence for the \thtp.

To avoid a possible dilution of a tiny signal by complex nuclear breakup processes involving a higher number of baryons, $p K_S^0$ candidates with a baryon like a $p$, $\bar{p}$, $d$ or $t$ passing the ($K_S^0$, $p$) vertex within a DOCA $<$ 3 mm have been rejected. In addition sub-samples with at least one associated non baryonic charged track are examined. The requirement of an associated $\pi^\pm$ or $K^\pm$ results in the observation of signals originating from $K^\ast(892)^+\rightarrow K_S^0 \pi^+$, $\bar{K}^\ast(892)^-\rightarrow K_S^0 \pi^-$, $\Lambda(1115)\rightarrow p\pi^-$, $\Lambda(1520)\rightarrow p K^-$ and indications of coupling to $a_0(980)^-$ and $\Lambda(1405)$.
In each case the corresponding $p K_S^0$ invariant-mass distribution shows no evidence for the \thtp.

\begin{figure}[htb]
\begin{center}
\epsfig{file=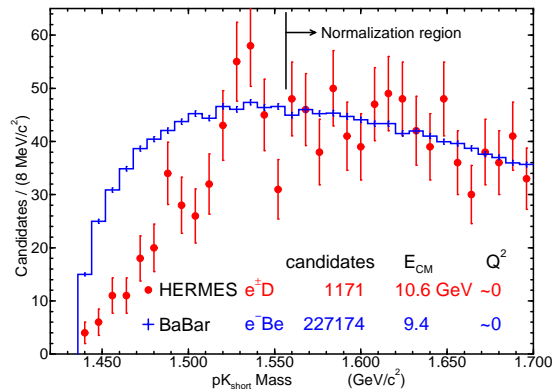, clip=, width=0.45\textwidth}
\caption{The HERMES $p K_S^0$ mass distribution \cite{HERMES} compared to the corresponding \babar\ distribution for electroproduction in Be normalized to the HERMES data for the region above 1.58\gevcc.}
\label{fig:hermes}
\end{center}
\end{figure}

\begin{figure}[htb]
\begin{center}
\epsfig{file=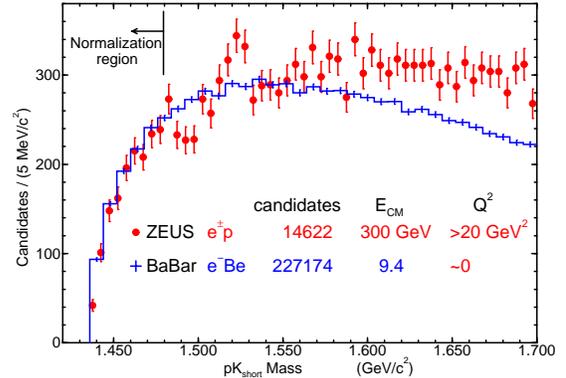, clip=, width=0.45\textwidth}
\caption{The ZEUS $p K_S^0$ mass distribution \cite{ZEUS} compared to the corresponding \babar\ distribution for electroproduction in Be normalized to the ZEUS data for the region below 1.48\gevcc.}
\label{fig:zeus}
\end{center}
\end{figure}

The study was then restricted solely to those regions which can be interpreted as corresponding to electro-production in the beampipe, which consists mostly of Be. Again no signal is seen.
Since there is no quantitative information on the flux of beam-halo electrons and positrons, it is not possible to derive upper limits for the production cross section or production rates as was done in section 3. Therefore the \babar\ electro-production results are compared to those of the HERMES \cite{HERMES} and ZEUS \cite{ZEUS} experiments which study $e^+ d$ and $e^\pm p$ reactions and thus have comparable experimental situations. Fig. \ref{fig:hermes} shows the superposition of the invariant-mass distributions measured by \babar\ and HERMES. The comparison seems to indicate a significant loss of acceptance for the HERMES experiment in the $p K_S^0$ mass region below $\approx 1.52$\gevcc resulting in a possible overestimation of the significance of the \thtp\ signal.
Compared to the mass distribution measured by the ZEUS experiment in Fig. \ref{fig:zeus}, the \babar\ spectrum shows no evidence for the poorly established baryon candidate $\Sigma(1480)$ \cite{PDG} and \thtp\ signal of the ZEUS analysis. This creates serious reservations about the significance of the \thtp\ observation claimed by this experiment.

\section{Conclusions}
In summary, we have performed a search for the reported pentaquark states \thtp, \ximm\ and \xiz\ in $e^+ e^-$ annihilations. We find large signals for known baryon states but no excess at the measured mass values for the pentaquark states. The measured upper limits for their production rates are 4-16 times lower than those expected for conventional baryons. These results do not directly disprove the reported positive evidence due to the unknown production mechanism of possible pentaquark states, but are nevertheless highly suggestive.

In addition we studied the production of \thtp\ in electro- and hadro-production events within the inner part of the \babar\ detector. Again no signal was observed. Furthermore, the comparison of the \babar\ results on electro-production in Be to those from the HERMES ($e^+ d$) and ZEUS ($e^\pm p$) experiments leads to the conclusion that the prior claims for the observation of \thtp\ in electro-production are less than convincing.

\end{document}